\documentstyle[11pt]{article}
\input amssym.def
\def\ra{\rangle}
\def\la{\langle}
\def\Nb{{\Bbb N}}

\def\Cb{{\Bbb C}}

\begin{document}
\baselineskip18pt
\begin{center}
{\LARGE\bf Separability of rank two quantum states on multiple
quantum spaces with different dimensions}
\end{center}
\vskip 2mm
\begin{center}
{\normalsize Shao-Ming Fei$^{1,\  2}$, Xiu-Hong Gao$^1$, Xiao-Hong
Wang$^1$, Zhi-Xi Wang$^1$, and Ke Wu$^1$}\\ {\small \sl $ ^1$
Department of Mathematics, Capital  Normal University, Beijing,
China.}\\ {\small \sl $ ^2$ Institute of Applied Mathematics,
University of Bonn,  53115 Bonn, Germany}
\end{center}

\vskip 2mm
\parbox{14cm}
{\footnotesize{\bf Abstract} We consider the separability of rank
two quantum states on multiple quantum spaces with different
dimensions. The sufficient and necessary conditions for
separability of these multiparty quantum states are explicitly
presented. A nonseparability inequality is also given, for the
case where one of the eigenvectors corresponding to nonzero
eigenvalues of the density matrix is maximally entangled.}
\vskip 4mm

Quantum entanglement is one of the most striking features of
quantum phenomena \cite{1}. It was first recognized by Schr\"odinger
\cite{5p} and Einstein, Podolsky and Rosen \cite{4p}, where a
description of the world called local realism was suggested. Bell
proved that the local realism implies constraints on the
predictions of spin correlations in the form of inequalities
(Bell's inequalities) \cite{Bell}. The feature of quantum
mechanics called nonlocality is one of the most apparent
manifestations of quantum entanglement. Nonlocality has been given
a lot of attention in foundational considerations, in the
discussion of Bell type inequalities and hidden variable models,
see e.g. \cite{ak}. Nonlocal correlations in quantum systems imply a kind of
entanglement among the quantum subsystems. The recent development
of quantum information theory showed that quantum entanglement can
have important practical applications (see e.g. \cite{8}). It is
playing very important roles in quantum information processing
such as quantum computation \cite{DiVincenzo}, quantum
teleportation \cite{teleport,teleport1,Ari00,teleport2} (for
experimental realization see \cite{telexp}), dense coding
\cite{dense} and quantum cryptographic schemes
\cite{crypto1,crypto2,crypto3}.

Due to interaction with environment, in real conditions one
encounters mixed states rather than pure ones. They can still
possess some residual entanglement. More specially, a mixed state
is considered to be entangled if it is not a mixture of product
states \cite{10}. In mixed states the quantum correlations are
weakened, hence the manifestations of mixed-state entanglement can
be very subtle \cite{10, 11, 12}. To investigate the structure of
mixed-state entanglement some beautiful works have been done in
quantifying entanglement \cite{17, 18, 19, 20, 21} for bipartite
systems and multipartite systems (see e.g. \cite{22,23}).
However most proposed measures of
entanglement for bipartite systems involve extremizations which
are difficult to handle analytically.
For multipartite systems, one even does not know how to define
the measures. Till now there is no general criterion that allows one to
distinguish whether a mixed state is separable or not.

The separability of pure states for bipartite systems is quite
well understood (cf. \cite{peresbook}). For mixed states, some
progress has been achieved in understanding the separability and
entanglement problem for bipartite systems (cf. \cite{primer}),
e.g., the proper definition of separable and entangled states
formulated by Werner \cite{10}, the Peres \cite{Peres}
criterion that all separable states necessarily have a positive
partial transpose (PPT), which is further shown to be also a sufficient
condition for separability in $2\times2$ and $2\times 3$ systems
\cite{ho96,tran}. Recently some new criterion are presented
in \cite{chenkai,rudolph}, which are necessary conditions for
a state to be separable and complement the
well-known PPT criterion in certain aspects.

An important property of entanglement is that the degree of
entanglement of two parts of a quantum system remains invariant
under local unitary transformations of these parts. Therefore the
invariants of local unitary transformations give rise to an
effective description of entanglement, especially for the study of
equivalence of quantum mixed states under local unitary
transformations, see e.g.,
\cite{Rains,Grassl,makhlin,linden,afpy}. In this paper, by using
the invariants of local unitary transformations, we study
sufficient and necessary conditions for (full) separability of
higher-dimensional quantum systems on $H_1\otimes
H_2\otimes\dots\otimes H_M$,  $\dim H_i=N_i,\ \ i=1,2,\cdots,M$,
which generalizes the results in \cite{4a} and \cite{20p}.
A general pure state on $H_1\otimes H_2\otimes\dots\otimes H_M$ is of the form
\begin{equation}
\label{16} |\Psi_M\ra=\sum_{k=1}^{M}\sum_{{i_k}=1}^{N_k} a_{i_1
i_2 \dots i_M}|e_{i_1}\ra\otimes |f_{i_2}\ra\otimes\dots\otimes |g_{i_M}\ra,
\hspace{0.6cm}a_{i_1 i_2\dots i_M}\in {\Bbb C}
\end{equation}
with $\sum a_{i_1 i_2\dots i_M}a_{i_1 i_2 \dots i_M}^\ast=1$
($\ast$ denoting complex conjugation) and $|e_{i_1}\ra,~
|f_{i_2}\ra,~...,~|g_{i_M}\ra$ the corresponding orthonormal basis of
complex Hilbert spaces $H_1,~H_2,~...,~H_M$.
$|\Psi_M\ra$ is said to be (fully) separable if $a_{i_1
i_2 \dots i_M}=a_{i_1}a_{i_2}...a_{i_M}$ for some
$a_{i_1},~a_{i_2},~...,~a_{i_M}\in\Cb$. A mixed state is described by a
density matrix $\rho$,
\begin{equation}
\label{rho1}
\rho=\sum_{i=1}^n p_i |\Psi_i\ra\langle\Psi_i\vert,
\end{equation}
where $n\in\Nb$, $0<p_i\leq 1$, $\sum_{i=1}^n p_i=1$,
$|\Psi_i\ra$s are pure states of the form (\ref{16}).
$\rho$ is said to be separable if there exits a decomposition (\ref{rho1})
such that all $|\Psi_i\ra$s are fully separable. We first present
an invariant of local unitary transformation which characterizes the
separability of pure states. Then we consider the separability
of density matrices with rank two. The
separability condition for these kinds of mixed states in arbitrary
dimensions is explicitly given. In addition, we present a
nonseparability inequality valid in the case where one of the
eigenvectors corresponding to nonzero eigenvalues of a density
matrix is maximally entangled.

We first consider the case of three different dimensional
quantum spaces. Let ${\cal H}_A$, ${\cal H}_B$ and ${\cal H}_C$ be
$M$, $N$ and $T$-dimensional complex
Hilbert spaces respectively. We denote $\{|e_i\ra\}_{i=1}^M$,
$\{|f_i\ra\}_{i=1}^N$ and $\{|g_i\ra\}_{i=1}^T$ the orthonormal
basis in ${\cal H}_A$, ${\cal H}_B$, and ${\cal H}_C$. A general
pure state on ${\cal H}_A\otimes {\cal H}_B\otimes {\cal H}_C$ is
of the form
\begin{equation}
\label{1} |\Psi\ra=\sum_{i=1}^M\sum_{j=1}^N \sum_{k=1}^T
a_{ijk}|e_i\ra\otimes |f_j\ra\otimes |g_k\ra, \ \  a_{ijk}\in {\bf C}
\end{equation}
with the normalization $\sum_{i=1}^M\sum_{j=1}^N\sum_{k=1}^T
a_{ijk}a_{ijk}^*=1$.

Let $U_1$, $U_2$ and $U_3$ denote unitary transformations on the
Hilbert space ${\cal H}_A$, ${\cal H}_B$ and ${\cal H}_C$,
respectively, such that
$$
U_1|e_i\ra=\sum_{j=1}^M b_{ij}|e_j\ra, \ \   U_2|f_i\ra=\sum_{j=1}^N c_{i j}|f_{j}\ra,
$$
$$
U_3|g_i\ra=\sum_{j=1}^T d_{i j}|g_{j}\ra,\ \ b_{ij},
\\ c_{i j}, \\d_{i j}\in {\bf C}
$$
and $\sum _{j=1}^M b_{ij}b_{kj}^*=\delta_{ik}$, $\sum _{j=1}^N
c_{ij}c_{kj}^*=\delta_{ik}$, $\sum _{j=1}^T
d_{ij}d_{kj}^*=\delta_{ik}$ (with $\delta_{ik}$ the usual
Kronecker's symbol ). We call a quantity an invariant associated
with the state $|\Psi\ra$ if it is invariant under all local
unitary transformations of $U_1\otimes U_2\otimes U_3$.
In \cite{5} we have presented a way in constructing invariants
by contracting the subindices in $a_{ijk}$ for multipartite
quantum systems with equal dimensions.
By generalizing the results in \cite{5},
the following quantities are straightforward shown to be
invariants under local unitary transformations:
$$
\begin{array}{ll} I_0=\displaystyle\sum_{i=1}^{M}\sum_{j=1}^{N}\sum
_{k=1}^T a_{ijk}a_{ijk}^*,& I_1=\displaystyle\sum_{i,p=1}^M\sum
_{j,q=1}^N \sum_{k,m=1}^T a_{ijk}a_{ijm}^*
a_{pqm}a_{pqk}^*,\\[3mm]
I_2=\displaystyle\sum_{i,p=1}^M\sum _{j,q=1}^N \sum_{k,m=1}^T
a_{ikj}a_{imj}^* a_{pmq}a_{pkq}^*,\hspace{0.2cm}&
I_3=\displaystyle\sum_{i,p=1}^M\sum _{j,q=1}^N \sum_{k,m=1}^T
a_{kij}a_{mij}^* a_{mpq}a_{kpq}^*.
\end{array}
$$
For the case of a pair of qubits, the quantity concurrence characterizes
the degree of entanglement \cite{HillWootters}. This concurrence has a
simple expression in terms of invariants \cite{5}. For high dimensional
multipartite quantum systems with equal dimensions, we defined a generalized
concurrence which has a similar expression in terms of invariants.
This generalized concurrence is generally no longer a measure of entanglement,
but it characterizes the separability and maximally entanglement of a quantum
state. For high dimensional multipartite quantum systems with different dimensions,
there is no Schmidt-like decompositions in general. Nevertheless we can still
define the following quantity to be a generalized concurrence,
\begin{equation} \label{2}
\begin{array}{crl}
C_{MNT}^3&=&C_{MNT}^3(|\Psi\ra)=
\displaystyle\sqrt{2(3I_0^2-I_1-I_2-I_3)}\\[4mm]
&=&\displaystyle\sqrt{\sum (|a_{ijk}a_{pqm}-
a_{ijm}a_{pqk}|^2
+|a_{ijk}a_{pqm}-a_{iqk}a_{pjm}|^2+|a_{ijk}a_{pqm}-a_{pjk}a_{iqm}|^2)}.
\end{array}
\end{equation}
Different from the the case in \cite{5}, $C_{MNT}^3(|\Psi\ra)=1$
does not mean that the quantum state $|\Psi\ra$ is maximally entangled.
But we still have

{\bf Lemma 1.} \ $C_{MNT}^3(|\Psi\ra)=0$ if and only if $|\Psi\ra$ is
separable.

{\bf Proof.} \  If $|\Psi\ra$ is factorizable, i.e., $a_{ijk}=a_i b_j c_k$,
for some $a_i, b_j, c_k \in {\Bbb C}$, it is obvious that $C_{MNT}^3=0$.

Conversely, because $|\Psi\ra\neq 0$, there exists
$p_0, q_0, m_0$ such that $a_{p_0q_0m_0}\neq 0$. If $C_{MNT}^3=0$,
from $|a_{ijk}a_{pqm}-a_{ijm}a_{pqk}|=0$ we have $a_{ijk}=a_{ij} b_k$,
for some $a_{ij}, b_k \in {\Bbb C}$. From the rest terms in (\ref{2}),
we further get $a_{ijk }=a_i'b_j' c_k'$, for some $a_i'$, $b_j'$,
$c_k' \in {\Bbb C}$, i.e., $|\Psi\ra$ is separable. $\Box$

Now let $\rho$ be a rank two state on ${\cal H}_A\otimes {\cal
H}_B\otimes {\cal H}_C$ , with $|E_1\ra$, $|E_2\ra$ being its two
orthonormal eigenvectors corresponding to the two nonzero eigenvalues:
\begin{equation}\label{3}
\rho=p|E_1\ra\la E_1|+q|E_2\ra\la E_2|,
\end{equation}
where $q=1-p\in (0, 1)$. Generally
\begin{equation}\label{es1}
|E_{s_1}\ra=\sum_{i=1}^M\sum_{j=1}^N\sum_{k=1}^T
a_{ijk}^{s_1}|e_i\ra\otimes |f_j\ra\otimes |g_k\ra,~~~
a_{ijk}^{s_1}\in {\Bbb C},~~~s_1=1,2
\end{equation}
with normalization
$\sum_{i=1}^M\sum_{j=1}^N\sum_{k=1}^T a_{ijk}^{s_1}
(a_{ijk}^{s_1})^*=1$.

Using Lemma 1, we have that $|\Psi\ra=\sum_{i=1}^M\sum_{j=1}^N
\sum_{k=1}^T a_{ijk}|e_i\ra\otimes |f_j\ra\otimes |g_k\ra$ is separable if and
only if $C_{MNT}^3=0$, i.e.,
\begin{equation}
\label{4} a_{ijk}a_{pqm}= a_{ijm}a_{pqk}, \hspace{0.2cm}
a_{ijk}a_{pqm}= a_{iqk}a_{pjm}, \hspace{0.2cm} a_{ijk}a_{pqm}=
a_{pjk}a_{iqm}, \hspace{0.2cm} \forall i, j, k, p, q, m.
\end{equation}

And a vector of the form $|E_1\ra+\lambda |E_2\ra$, $\lambda \in
{\Bbb C}$, is separable if and only if
$C_{MNT}^3(|E_1\ra+\lambda |E_2\ra)=0$. Using  (\ref{es1}) we have that
$\lambda$ is a common root of the following equation set $Eq_s^I$:
\begin{equation}
\label{6}
\alpha_s^I
\lambda^2+\beta_s^I\lambda+\gamma_s^I=0,~~~~~~s=1,~2,~3,
\end{equation}
where
\begin{equation}\label{5}
\begin{array}{l}
\alpha_1^{I}=a_{ijk}^2 a_{pqm}^2-a_{ijm}^2 a_{pqk}^2,
\hspace{0.2cm} \alpha_2^{I}=a_{ijk}^2 a_{pqm}^2-a_{iqk}^2
a_{pjm}^2, \hspace{0.2cm} \alpha_3^{I}=a_{ijk}^2
a_{pqm}^2-a_{pjk}^2 a_{iqm}^2, \\[3mm]
\beta_1^{I}=a_{ijk}^2 a_{pqm}^1+a_{ijk}^1
a_{pqm}^2-a_{ijm}^2 a_{pqk}^1-a_{ijm}^1 a_{pqk}^2,
\hspace{0.2cm}\\[3mm] \beta_2^{I}=a_{ijk}^2
a_{pqm}^1+a_{ijk}^1
a_{pqm}^2-a_{iqk}^2 a_{pjm}^1-a_{iqk}^1 a_{pjm}^2,\\[3mm]
\beta_3^{I}=a_{ijk}^2 a_{pqm}^1+a_{ijk}^1
a_{pqm}^2-a_{pjk}^2 a_{iqm}^1-a_{pjk}^1 a_{iqm}^2,\\[3mm]
\gamma_1^{I}=a_{ijk}^1 a_{pqm}^1-a_{ijm}^1
a_{pqk}^1,\hspace{0.2cm} \gamma_2^{I}=a_{ijk}^1
a_{pqm}^1-a_{iqk}^1 a_{pjm}^1, \hspace{0.2cm}
\gamma_3^{I}=a_{ijk}^1 a_{pqm}^1-a_{pjk}^1 a_{iqm}^1,
\end{array}
\end{equation}
$I=\{i, j, k, p, q, m\}$, $\forall i,p\in
\{1,2,\dots,M\}$, $j,q\in \{1,2,\dots,N\}$ and $k,m\in \{1,2,\dots,T\}$.

{\bf Lemma 2.}\ If $|E_2\ra$ is not separable, then $\rho$ is
separable if and only if (\ref{6}) have two distinct roots.

{\bf Proof.} \  Suppose $\rho$ has a decomposition,
$\rho=\sum_{t=1}^l p_t' |U_t\ra\la U_t|$,
with $l$ some positive integer and  $0< p_t'<1, \ \sum p_t'=1,\
|U_t\ra$ being separable, $\forall t$. We can write them as linear
combinations of the two eigenvectors $|E_1\ra$  and $|E_2\ra$ which
span the range of $\rho:\  |U_t\ra=c_1^t |E_1\ra+c_2^t |E_2\ra$ (for
some $c_1^t,c_2^t \in {\Bbb C}$). As $|U_t\ra\neq 0, c_1^t, c_2^t$
can not be all 0. Without losing generality, let $c_1^t\neq 0$.
$|U_t\ra$  is then of the form $|E_1\ra+\lambda_t |E_2\ra,
\lambda_t=c_2^t/c_1^t$.
From Lemma 1 we have that $|U_t\ra$ is separable if and only if
parameter $\lambda_t$ is a common root of the corresponding
equation set $Eq_s^I:~\alpha_s^I \lambda^2+\beta_s^I
\lambda+\gamma_s^I=0$.

Because $|E_2\ra$ is not separable, not all $\lambda_t's$ can be
equal, otherwise all the $U_t's$ would be constant multiples of a
fixed vector, and $\rho$ would be rank 1. On the other hand, as
$|E_2\ra$ is not separable, then $C_{MNT}^3$ is not zero. Hence
there is some $I_0, s_0$ such that $\alpha_{s_0}^{I_0}\neq 0$,
i.e., the relation $Eq_s^I$ is indeed a quadratic equation. It
must have exactly two roots, and so there are two values that are
the only possible choices for the $\lambda_t's$. But in order that
there is not only one possible choice, the above two roots must be
different.  And all the relations $Eq_s^I$ have these two
different roots.

Let $\mu_1,\mu_2$ be two distinct roots, which are common to all
of the equations $Eq_s^I$. Each vector  $|U_t\ra$ is either of the
form $|E_1'\ra=(|E_1\ra+\mu_1 |E_2\ra)/\sqrt{1+|\mu_1|^2}$, or of
the form $|E_2'\ra=(|E_1\ra+\mu_2 |E_2\ra)/\sqrt{1+|\mu_2|^2}$.

Therefore we can write $\rho$ as $\rho=p'|E_1'\ra\la
E_1'|+(1-p')|E_2'\ra\la E_2'|$ with $0< p'<1$. Comparing the
coefficients of $|E_k\ra\la E_l|$, $k,l=1,2$, with the ones in the
expression (\ref{3}), we get the following two relations:
\begin{equation}
\label{7} \frac{p'}{1+|\mu_1|^2}+\frac{1-p'}{1+|\mu_2|^2}=p,
~~~~~\frac{\mu_1p'}{1+|\mu_1|^2}+\frac{\mu_2
(1-p')}{1+|\mu_2|^2}=0.
\end{equation}
Solving the above equations for $p$ and $p'$ we get
\begin{equation}
\label{9}
p=(1-\mu_1\mu_2\frac{z^\ast}{z})^{-1}, \hspace{0.7cm}
p'=\frac{\mu_2(1+|\mu_1|^2)}{z-\mu_1\mu_2z^\ast},
\end{equation}
where $z=\mu_2-\mu_1$.

Conversely, let $\mu_1, \mu_2$ be two distinct roots, which are
common to all of the equations $Eq_s^I$. From above discussion we
have $\rho=p'|E_1'\ra\la E_1'|+(1-p')|E_2'\ra\la E_2'|$, i.e.,
$\rho$ is separable. $\Box$

Before getting the separability criterion for general $\rho$ in (\ref{3}),
we first deal with the case where the coefficients
$a_{ijk}^{s_1}$ in (\ref{es1}) are all real. We
will use the following result:

{\bf Lemma 3.} \  For a quadratic equation $a x^2+bx+c=0$ with
$a,b,c\in {\Bbb R}$, $a\neq 0$, and roots $\alpha$, $\beta$ with
$\alpha\neq \beta$,
$\gamma=(\alpha^\ast-\beta^\ast)/(\alpha-\beta)$  is either $1$
or $-1$.

{\bf Theorem 1.}\ If all $a_{ijk}^{s_1}$ are real, $\rho$ is
separable if and only if one of the following
quantities($\triangle_1$
 or $\triangle_2$) is zero:
\begin{equation}
\label{10}
\triangle_1=\sum
|\gamma_s^I-(1-p^{-1})\alpha_s^I|^2+\sum
|\beta_s^I\alpha_{s'}^{I'}-\alpha_s^I\beta_{s'}^{I'}|^2,
\end{equation}
\begin{equation}
\label{11}
\triangle_2=\sum
|\gamma_s^I+(1-p^{-1})\alpha_s^I|^2+\sum |\beta_s^I|^2,
\end{equation}
or, equivalently, one of the following two sets of
relations (\ref{12}) or (\ref{13}) hold:
\begin{equation}\label{12}
\gamma_s^I=(1-p^{-1})\alpha_s^I,
\hspace{0.6cm}
\beta_s^I\alpha_{s'}^{I'}=\alpha_s^I\beta_{s'}^{I'}
\end{equation}
\begin{equation}\label{13}
\gamma_s^I=-(1-p^{-1})\alpha_s^I,
\hspace{0.6cm}
\beta_s^I=0
\end{equation}
where $s, s'=1, 2, 3$, and $I, I'=\{i,j,k,p,q,m\}$,
$\forall i,p\in \{1,2,\dots,M\}$,
$j,q\in \{1,2,\dots,N\}$, and $k,m\in \{1,2,\dots,T\}$.

{\bf Proof.}\  We prove the necessity part of the theorem in two
cases:

Case 1. $|E_2\ra$ is not separable.

$a).$ We get that (\ref{6}) have
two distinct roots from Lemma 2. These two roots are the solutions
to all the relations $Eq_s^I$. Consider for any  $ s=1,2,3$, $I=\{i,
j, k, p, q, m\}$, $\forall i,p\in \{1,2,\dots,M\}$, $
j,q\in \{1,2,\dots,N\}$, and $k,m\in \{1,2,\dots,T\}$.

If $\alpha_s^I\neq 0$, the corresponding relation (\ref{6}) is not an
identity. All the quadratic equations in the set $Eq_s^I$ have the
same two distinct roots. From the standard theory of quadratic
equations, we have
\begin{equation}
\label{12p}
\beta_s^I\alpha_{s_0}^{I_0}=\beta_{s_0}^{I_0}\alpha_s^I,~~~~~~
\gamma_s^I\alpha_{s_0}^{I_0}=\gamma_{s_0}^{I_0}\alpha_s^I.
\end{equation}

If $\alpha_s^I =0$, then the equations $Eq_s^I$ become
identities, i.e.,  $\beta_s^I$ and $\gamma_s^I$ must be $0$ too,
because otherwise at least one of the relations $Eq_s^I$ would be
a linear equation, and there would be no two distinct roots. Thus
in this case (\ref{12p}) also hold.

$b).$  Because all $a_{ijk}^{s_1}$ are real number,  $\mu_1$ and
$\mu_2$ are roots of a quadratic equation with real coefficients.
From Lemma 3, $\mu_1\mu_2 =1-p^{-1}$ or $-(1-p^{-1})$. Since
$\mu_1\mu_2$ is real, the solution for $p'$ in (\ref{9}) implies that
$\mu_2/(\mu_2-\mu_1)$ is real, which is possible if and only if
either the roots are both real or the roots are both purely
imaginary. In the first case, let $\mu_2> \mu_1$, we have
$\mu_1\mu_2=1-p^{-1}$. From (\ref{9}), we get the condition that
$p'\in[0,1]$, which is equivalent to $\mu_2>0$, $\mu_1<0$. In  the
second case, we have $\mu_1\mu_2=-(1-p^{-1})$. The condition for
having purely imaginary roots of quadratic equations gives that
$\beta_s^I=0, \forall I$ and $s$.

$c).$ Finally, we observe that $\mu_1\mu_2$ is nothing but the
ratio $\gamma_{s_0}^{I_0}/ \alpha_{s_0}^{I_0}$, which is either
$1-p^{-1}$ or $-(1-p^{-1})$. Therefore  we conclude that either
$\gamma_s^I=(1-p^{-1})\alpha_s^I$ or
$\gamma_s^I=-(1-p^{-1})\alpha_s^I$ for any $I$ and $s=1,2,3$.
Relation (\ref{10}), (\ref{11}) are verified.

Case 2. $|E_2\ra$ is separable.

In this case from (\ref{4}), we have
$\alpha_s^I=0$, $\forall I$ and $s$. Since not all of the
$|U_t\ra$ can be multiples of $|E_2\ra$, we must have at least one
choice of $\lambda$ such that $|E_1\ra+\lambda |E_2\ra$ is
separable. This must be a common root to all equations $Eq_s^I$ as
before. All these equations are now linear ones. When all
$\beta_s^I=\gamma_s^I=0$, it is easy to see that $|E_1\ra$ is
separable. Excluding this case, we see that there is only one
 possible choice of $\lambda$. Then $\rho$ can be expressed as
$$\rho=p''|E_2\ra\la E_2|+(1-p'')\frac{|E_1+\lambda E_2\ra\la
E_1+\lambda E_2|}{1+|\lambda|^2}.$$ That is $p''=1$, which is a
contradiction. Thus, if $|E_2\ra$ is separable, $|E_1\ra$ must be
separable too. It is clear that in this case (\ref{10}) and (\ref{11}) hold.

Now we prove the sufficiency part for the theorem. If (\ref{10}) or (\ref{11})
holds, it is clear the equations $Eq_s^I$ have common roots. If
$|E_2\ra$ is not separable, then not all of these equations are
identities. And there are at most two common roots. If (\ref{10}) holds,
the product of the two roots must be $1-p^{-1}<0$, so that the two
roots are real and unequal. If (\ref{11}) holds, the two roots must be
purely imaginary. So in these two cases, we get that $\rho$ is
separable in terms of Lemma 2. If $|E_2\ra$ is separable, from
(\ref{10}) or (\ref{11}) we know $|E_1\ra$ is separable too and $\rho$ is
separable. $\Box$

Generalizing the results in Theorem 1, we have, for the complex
$a_{ijk}^{s_1}$,

{\bf Theorem 2.}\ $\rho$ is separable if and only if there is
$\theta\in {\Bbb R}$ such that
\begin{equation}
\label{14}
\begin{array}{l}
\gamma_s^I=e^{i\theta}(1-p^{-1})\alpha_s^I, \hspace{0.6cm}
\beta_s^I\alpha_{s'}^{I'}=\alpha_s^I\beta_{s'}^{I'},
\end{array}
\end{equation}
where $s, s'=1,2,3$, $I, I'=\{i,j,k,p,q,m\}$, $\forall i,p\in
\{1,2,\dots,M\}$, $j,q\in \{1,2,\dots,N\}$ and $k,m\in \{1,2,\dots,T\}$,
and
\begin{equation}
\label{15} \frac{\mu_2(1+|\mu_1|^2)}{z-\mu_1\mu_2z^\ast}\in
[0,1],
\end{equation}
where $z=e^{i\theta}z^\ast, z=\mu_2-\mu_1\neq 0$, $\mu_1$ and
$\mu_2$ are the roots of the equation
$\alpha_s^I\lambda^2+\beta_s^I\lambda+\gamma_s^I=0$ for some $I$
and $s$ such that $\alpha_s^I\neq 0$.

{\bf  Proof.}\  The proof of necessity is similar to the proof of
the corresponding part in Theorem 1. One only needs to note that
since $z/z^\ast$ is of modulus 1, a phase factor $e^{i\theta}$
appears in this case.

Now if (\ref{14}) holds, it is clear that the equations $Eq_s^I$
have common roots. If $|E_2\ra$ is not separable, then  some of
the $\alpha_s^I$ are nonzero. The corresponding equations $Eq_s^I$
have exactly two different roots by condition (\ref{15}).
Therefore $\rho$ is separable from Lemma 2. If $|E_2\ra$ is
separable, by (\ref{14}) we know that all $\gamma_s^I$ are $0$.
Hence both $|E_2\ra$ and $|E_1\ra$ are separable, and so is
$\rho$. $\Box$

{\bf Corollary 1.}\  Let $|E_2\ra$ be the maximally entangled state
given by $|E_2\ra=(1/\sqrt{N_1})\sum_{i=1}^{N_1} |e_i\ra\otimes
|f_i\ra\otimes |g_i\ra$, where $N_1=min\{M,N,T\}$. For any vector $|E_1\ra$
which is orthogonal to $|E_2\ra$, $\rho=p|E_1\ra \la E_1|+(1-p)
|E_2\ra \la E_2|$ is not separable for $0<p<1/2$.

{\bf Proof.} \  Let
$$
C_{(1)}=\sqrt{\sum_{I,s}|\gamma_s^I|^2}
\hspace{0.2cm} \ {\rm and } \hspace{0.2cm}
C_{(2)}=\sqrt{\sum_{I,s}|\alpha_s^I|^2}
$$
be the generalized concurrences associated with the states $|E_1\ra$
and $|E_2\ra$, respectively, where $s=1,2,3$, $I=\{i,j,k,p,q,m\}$,
$\forall i,p\in \{1,2, \dots, M\}$, $j,q\in \{1,2, \dots,
N\}$ and $k,m\in \{1,2, \dots,T\}$.

Suppose $\rho$ is separable. As a pure
state is separable if and only if the corresponding
generalized concurrence is zero, a necessary condition for
separability is that $C_{(1)}$ and $C_{(2)}$
should be inversely proportional
to the contribution of the corresponding pure state to $\rho$, i.e. the
eigenvalue corresponding to that eigenstate.
From the separable condition
$\gamma_s^I=e^{i\theta}(1-p^{-1})\alpha_s^I$, we
get $C_{(1)}=\displaystyle\frac{1-p}{p}C_{(2)}$. As $|E_2\ra$ is
maximally entangled, $C_{(2)}\neq 0$, and
$C_{(1)}/C_{(2)}=\displaystyle\frac{1-p}{p}\leq 1$, so we have
$p\geq 1/2$, which is a contradiction. $\Box$

The above approach can be extended to the case of multiquantum
systems. We consider now the separability of $|\Psi_M\ra$ given by
(\ref{16}). We have a quadratic $I_0=\sum a_{i_1 i_2 \dots i_M}a_{i_1 i_2 \dots
i_M}^*$ and $d=2^{M-1}-1$ biquadratic invariants:
$$
I_{TS}=\sum a_{TS}a_{TS'}^*a_{T'S'}a_{T'S}^*,
$$
where $T$ and $T'$ are all possible  nontrivial subset of
$I=\{i_1,i_2,\dots, i_M \}$, $I'=\{\tilde{i}_1, \tilde{i}_2, \dots,
\tilde{i}_M\}$, respectively, $\forall i _k,
\tilde{i}_k=1,2,\cdots, N_k$, $k=1, 2, \dots, M$, i.e.,
\begin{equation}\label{TS}
T\neq\emptyset,~~ T\neq I,~~ S=I\backslash T;~~~~~
T'\neq\emptyset,~~ T'\neq I',~~ S'=I'\backslash T'.\\
\end{equation}

$T$ and $T'$ are subindices of $a$, associated with the same
Hilbert spaces. A generalized concurrence can be defined by
\begin{equation}
\label{17} C_{N_1,N_2,\dots,N_M}^M
=\sqrt{2(dI_0^2-I_1-I_2-\cdots-I_d)}=\sqrt{\sum_p
|a_{TS}a_{T'S'} -a_{TS'}a_{T'S}|^2},
\end{equation}
where $\sum_p$ stands for the summation over all possible
combination of the indices of $T$ and $S$.
Similar to Lemma 1, one can prove:

{\bf Lemma 4.}\  $C_{N_1,N_2,\dots,N_M}^M=0$ if and only if
$|\Psi_M\ra$ is separable.

Let $\rho$ be a rank two state on $H_1\otimes
H_2\otimes\cdots\otimes H_M$, with $|E_1\ra, |E_2\ra$ being its
two orthonormal eigenvectors corresponding to the two nonzero
eigenvalues:
\begin{equation}
\label{18}
\begin{array}{l}
\rho=p|E_1\ra\la E_1|+q|E_2\ra\la E_2|,
\end{array}
\end{equation}
where $q=1-p\in (0,1)$. Generally
\begin{equation}\label{ees}
|E_{s_1}\ra=
\sum_{k=1}^{M}\sum_{{i_k}=1}^{N_k} a_{i_1 i_2 \dots
i_M}^{s_1}|e_{i_1}\ra\otimes |f_{i_2}\ra\otimes\dots\otimes |g_{i_M}\ra,
~~~a_{i_1i_2\dots i_M}^{s_1} \in {\Bbb C},~~~s_1=1,2
\end{equation}
with normalization  $\sum a_{i_1 i_2 \dots i_M}^{s_1} (a_{i_1 i_2 \dots
i_M}^{s_1})^*=1$.

Using Lemma 4, we have  $|\Psi_M\ra$ is separable if and only if
\begin{equation}
\label{19}
\begin{array}{l}
a_{TS}a_{T'S'}=a_{TS'}a_{T'S},
\end{array}
\end{equation}
where $T$ (resp. $T'$) are all possible nontrivial subset of
$I=\{i_1,i_2,\dots, i_M \}$ (resp. $I'=\{\tilde{i}_1, \tilde{i}_2, \dots,
\tilde{i}_M\}$) in the sense of (\ref{TS}).

With the notations:
$$
\begin{array}{l}
\alpha_{TS}^{T'S'}=a_{TS}^2 a_{T'S'}^2-a_{TS'}^2 a_{T'S}^2,
\hspace{0.2cm}
\gamma_{TS}^{T'S'}=a_{TS}^1 a_{T'S'}^1-a_{TS'}^1 a_{T'S}^1,\\[3mm]
\beta_{TS}^{T'S'}=a_{TS}^2 a_{T'S'}^1+a_{TS}^1
a_{T'S'}^2-a_{TS'}^2 a_{T'S}^1-a_{TS'}^1 a_{T'S}^2,
\end{array}
$$
we have that a vector of the form $|E_1\ra+\lambda |E_2\ra,
\lambda \in {\Bbb C}$, is separable if and only if   $\lambda$ is
a common root of the following equation set:
\begin{equation}
\label{20}
\begin{array}{l}
Eq_{TS}^{T'S'}:\
\alpha_{TS}^{T'S'}\lambda^2+\beta_{TS}^{T'S'}\lambda+\gamma_{TS}^{T'S'}=0.
\end{array}
\end{equation}
Similar to the case of three  quantum  system, one has:

{\bf Lemma 5.}\ $\rho$ is separable if and only if (\ref{20}) have two
distinct roots.

From Lemma 4 and Lemma 5 it is straightforward to prove the
following conclusion:

{\bf Theorem 3.}\ If all $a_{i_1i_2\dots i_M}^{s_1}$ are real,
$\rho$ is separable if and only if one of the following quantities
($\triangle_1$ or $\triangle_2$) is zero:
\begin{equation}
\label{21} \triangle_1=\sum
|\gamma_{TS}^{T'S'}-(1-p^{-1})\alpha_{TS}^{T'S'}|^2+ \sum
|\beta_{TS}^{T'S'}\alpha_{T_1S_1}^{T_1'S_1'}
-\alpha_{TS}^{T'S'}\beta_{T_1S_1}^{T_1'S_1'}|^2,
\end{equation}
\begin{equation}
\label{22} \triangle_2=\sum
|\gamma_{TS}^{T'S'}+(1-p^{-1})\alpha_{TS}^{T'S'}|^2+\sum
|\beta_{TS}^{T'S'}|^2,
\end{equation}
or, equivalently, one of the following two sets of relations (\ref{23})
or (\ref{24}) hold:
\begin{equation}
\label{23} \gamma_{TS}^{T'S'}=(1-p^{-1})\alpha_{TS}^{T'S'},
\hspace{0.6cm}
\beta_{TS}^{T'S'}\alpha_{T_1S_1}^{T_1'S_1'}
=\alpha_{TS}^{T'S'}\beta_{T_1S_1}^{T_1'S_1'},
\end{equation}
\begin{equation}
\label{24} \gamma_{TS}^{T'S'}=-(1-p^{-1})\alpha_{TS}^{T'S'},
\hspace{0.6cm}
\beta_{TS}^{T'S'}=0,
\end{equation}
where $T$ (resp. $T'$) are all possible  nontrivial subset of
$I=\{i_1,i_2,\dots, i_M \}$, (resp. $I'=\{\tilde{i}_1,
\tilde{i}_2, \dots, \tilde{i}_M\}$), $S=I\backslash T$,
$S'=I'\backslash T'$. $T_1$ (resp. $T_1'$) are all possible
nontrivial subset of $J=\{j_1,j_2,\dots, j_M \}$ (resp.
$J'=\{\tilde{j}_1, \tilde{j}_2, \dots, \tilde{j}_M\}$),
$S_1=J_1\backslash T_1$, $S_1'=J_1'\backslash T_1'$.

Extend Theorem 3 to general complex coefficients $a_{i_1i_2\dots
i_M}^{s_1}$, we have

{\bf Theorem 4.}\ $\rho$ is separable if and only if there is
$\theta\in {\Bbb R}$ such that
 \begin{equation}
\label{25}
\gamma_{TS}^{T'S'}=e^{i\theta}(1-p^{-1})\alpha_{TS}^{T'S'},
\hspace{0.6cm}
 \beta_{TS}^{T'S'}\alpha_{T_1S_1}^{T_1'S_1'}=
 \alpha_{TS}^{T'S'}\beta_{T_1S_1}^{T_1'S_1'}
\end{equation}
\begin{equation}
\label{26} \frac{\mu_2(1+|\mu_1|^2)}{z-\mu_1\mu_2z^\ast}\in
[0,1].
\end{equation}
where $T,T',S,S',T_1,T_1',S_1,S_1'$ are defined  as in Theorem 3,
$z=e^{i\theta}z^\ast, z=\mu_2-\mu_1\neq 0$, $\mu_1$ and $\mu_2$
are the roots of the equation
$\alpha_{TS}^{T'S'}\lambda^2+\beta_{TS}^{T'S'}\lambda+\gamma_{TS}^{T'S'}=0$
for some $ T, S,T', S'$ such that $\alpha_{TS}^{T'S'}\neq 0$.

The criterion is operational.
For a given rank two density matrix  on $H_1\otimes
H_2\otimes\cdots\otimes H_M$, to find its separability one only
needs to calculate the two eigenvectors $|E_1\ra$, $|E_2\ra$
corresponding to the two nonzero eigenvalues and check if formula
(\ref{25}) is satisfied or not. The (finite) number of equations
need to be checked depends on the dimensions of $H_i$ and $M$. For given
dimensions, it increases according to $2^{M-1}-1$.

{\bf Corollary 2.} \  Let $|E_2\ra$ be the maximally entangled
vector given by $|E_2\ra=(1/\sqrt{N})\sum_{i=1}^N |e_{i}\ra \otimes
|f_{i}\ra\otimes \cdots \otimes |g_{i}\ra$, where
$N=min\{{N_1,N_2,\dots,N_M}\}$. For any vector $|E_1\ra$ which is
orthogonal to $|E_2\ra$, $\rho =p|E_1\ra\la E_1|+(1-p)|E_2\ra\la
E_2|$ is not separable for $0<p<1/2$.

We have studied the sufficient and necessary conditions for
separability of rank two mixed states in higher-dimensional multipartite
quantum systems on $H_1\otimes H_2\otimes\dots\otimes H_M$. The
separability condition for these kind of mixed states in arbitrary
dimensions is explicitly presented, together with a nonseparability inequality
for the case where one of the eigenvectors
corresponding to nonzero eigenvalues of a density matrix is
maximally entangled.

\bigskip
\noindent{\bf Acknowledgement.}\ This work has been supported by
NSF of China (No. 19975061) and the National Key Project for Basic
Research of China (G1998030601). We would like to thank the referee
for very useful comments.


\vskip 8mm
\begin{thebibliography}{20}

\bibitem{1} A. Peres, Quantum Mechanics: Concepts and Methods, Kluwer, Dordrecht 
(1993).

\bibitem{5p} E. Schrodinger: Naturwissenschaften 23, 807 (1935).

\bibitem{4p} A. Einstein, B. Podolsky and N. Rosen: Phys. Rev. 47, 777 (1935).

\bibitem{Bell}
J. Bell, {\em Physics } 1, 195 (1964).

\bibitem{ak}
{\it Quantum Theory: Reconsideration of Foundations}, edited by A.
Khrennikov, V\"axj\"o University Press 2002.

\bibitem{8} Phys. World, March 1998; J. Gruska Quantum Computing, McGraw-Hill,
London (1999).

\bibitem{DiVincenzo} See, for example, D.P. DiVincenzo,
{\em Science} 270, 255 (1995).

\bibitem{teleport} C.H. Bennett, G. Brassard, C. Cr\'epeau,
       R. Jozsa, A. Peres,
       and W.K. Wootters, {\em Phys. Rev. Lett.} 70, 1895 (1993).

\bibitem{teleport1} S. Albeverio and S.M. Fei,
          {\em Phys. Lett.} A 276, 8-11 (2000).

\bibitem{Ari00} G.M. D'Ariano, P.Lo Presti, M.F. Sacchi, {\it Phys. Lett.}
A 272, 32 (2000).

\bibitem{teleport2}
S. Albeverio, S.M. Fei and W.L. Yang, {\em Commun. Theor.
Phys.} 38, 301-304 (2002); {\em Phys. Rev.} A 66, 012301 (2002).

\bibitem{telexp}
D. Bouwmeester, J.-W. Pan, K. Mattle, M.
Elbl, H. Weinfurter and A. Zeilinger, Nature (London) 390, 575 (1997);\\
D. Boschi, S. Brance,
F. De Martini, L. Hardy and S. Popescu, Phys. Rev. Lett. 80, 1121 (1998);\\
A. Furusawa, J.L. S\o{}rensen, S. L. Braunstein, C. A. Fuchs, H. J.
Kimble and
E. S. Polzik: Science 282, 706 (1998);\\
M. A. Nielsen, E. Knill and R. Laflamme: Nature 396, 52 (1998).

\bibitem{dense} C.H. Bennett and S.J. Wiesner,
        {\em Phys. Rev. Lett.} 69, 2881 (1992).

\bibitem{crypto1}
A.~Ekert, {\em Phys. Rev. Lett.} 67, 661 (1991).

\bibitem{crypto2} D.~Deutsch, A.~Ekert, R. Jozsa, C. Macchiavello,
S. Popescu and A. Sanpera, {\em Phys. Rev. Lett.} 77, 2818 (1996).

\bibitem{crypto3} C.A. Fuchs, N. Gisin,
         R.B. Griffiths, C-S. Niu and
         A. Peres, {\em Phys. Rev.} A 56, 1163 (1997).

\bibitem{10} R. F. Werner, Phys. Rev. A 40, 4277 (1989).

\bibitem{11} S. Popescu, Phys. Rev. Lett. 72, 797 (1994).

\bibitem{12} S. Popescu, Phys. Rev. Lett. 74, 2619 (1995).

\bibitem{17} C. H. Bennett, D. P. DiVincenzo, J. Smolin and
W. K. Wootters: Phys. Rev. A 54, 3824 (1996).

\bibitem{18} V. Vedral, M. B. Plenio, M. A. Rippin and
P. L. Knight: Phys. Rev. Lett. 78, 2275 (1997).

\bibitem{19} V. Vedral and M. Plenio: Phys. Rev. A 57, 1619 (1998).

\bibitem{20} G. Vidal and R. Tarrach: Phys. Rev. A 59, 141 (1999).

\bibitem{21} G. Vidal: J. Mod. Opt. 47, 355 (2000).

\bibitem{22} M. Murao, M. B. Plenio, S. Popescu, V. Vedral and P.L. Knight: Phys. Rev.
A 57, R4075
(1998); W. D\"ur, J. I. Cirac and R. Tarrach: Phys. Rev. Lett. 83,
3562 (1999); N. Linden, S. Popescu and A. Sudbery: ibid. 83, 243 (1999).

\bibitem{23} C. H. Bennett, D. DiVincenzo, T. Mor, P. Shor, J. Smolin and B. Terhal:
Phys. Rev. Lett. 82, 5385 (1999).

\bibitem{peresbook} A. Peres, ``Quantum Theory: Concepts and Methods'',
Kluwer Academic Publishers (1995).

\bibitem{primer}
see M. Horodecki, P. Horodecki and R. Horodecki in ``Quantum
Information - Basic Concepts and Experiments'', Eds. G. Alber and
M. Weiner, (Springer, Berlin, 2000).\\
M. Lewenstein, D. Bru{\ss}, J. I. Cirac, B. Kraus, M. Ku\'s, J.
Samsonowicz, A. Sanpera and R. Tarrach, J. Mod. Phys. 47, 2481 (2000).

\bibitem{Peres} A. Peres  Phys. Rev. Lett. 77, 1413 (1996).

\bibitem{ho96} M. Horodecki, P. Horodecki and R. Horodecki,
Phys. Lett. A 223, 8 (1996).

\bibitem{tran} P. Horodecki, Phys. Lett. A 232, 333 (1997).

\bibitem{chenkai}
K. Chen and L.A. Wu, Phys. Lett. A 306, 14 (2002).\\
K. Chen and L.A. Wu, {\it The Generalized Partial Transposition
Criterion for Separability of Multipartite Quantum States},
quant-ph/0208058.

\bibitem{rudolph}
O. Rudolph, {\it A separability criterion for density operators},
J. Phys. A: Math. Gen. 33, 3951-3955 (2000).\\
O. Rudolph, {\it Further results on the cross norm criterion for separability},
quant-ph/0202121.\\
O. Rudolph, {\it Some Properties of the Computable Cross Norm Criterion for
Separability}, quant-ph/0212047.

\bibitem{Rains}
E.M.~Rains, {\em IEEE Transactions on Information Theory} 46, 54-59 (2000).

\bibitem{Grassl}
M.~Grassl, M.~R\"otteler and T.~Beth, {\em Phys. Rev. A} 58, 1833 (1998).

\bibitem{makhlin}
Y. Makhlin, {\it Nonlocal properties of two-qubit gates and mixed
states and optimization of quantum computations},
quant-ph/0002045.

\bibitem{linden}
N. Linden, S. Popescu and A. Sudbery, {\em Phys. Rev. Lett.} 83, 243 (1999).

\bibitem{afpy}
S. Albeverio, S.M. Fei, P. Parashar and W.L. Yang,
{\it Nonlocal Properties and Local Invariants for Bipartite Systems},
MPI 02-73, 2002.

\bibitem{4a} S. Albeverio, S.M. Fei and D. Goswami,
Phys. Lett. A 286, 91-96 (2001).

\bibitem{20p} S.M. Fei, X.H. Gao, X.H. Wang, Z.X. Wang and K. Wu,
Phys. Lett. A 300, 559-566 (2002).

\bibitem{5} S. Albeverio and S.M. Fei, J. Opt. B: Quantum Semiclass. Opt. 3, 223 (2001).

\bibitem{HillWootters} S.~Hill and W.K.~Wootters, {\em Phys. Rev. Lett.}
78, 5022 (1997).\\
W.K.~Wootters, {\em Phys. Rev. Lett.} 80, 2245 (1998).

\end{thebibliography}
\end{document}